\documentclass[11pt, journal, draftclsnofoot, onecolumn]{IEEEtran}

\usepackage{mathtools,enumitem,color,subfigure,bbm}
\usepackage{graphicx,tabularx,array}
\usepackage{hyperref}

\usepackage{amsthm,mathrsfs}
\usepackage{makecell}
\usepackage{multirow}
\usepackage[T1]{fontenc}
\usepackage{cite}
\usepackage{txfonts}

\IEEEoverridecommandlockouts

\definecolor{green}{rgb}{0.6627,0.8196,0.5568}
\definecolor{purple}{rgb}{ 0.7647,    0.6078,    0.8824}

\DeclareMathOperator*{\argmin}{arg\,min}

\newcommand{\discardpages}[1]{
  \xdef\discard@pages{#1}
  \AtBeginShipout{
    \renewcommand*{\do}[1]{
      \ifnum\value{page}=##1\relax%
        \AtBeginShipoutDiscard
        \gdef\do####1{}
      \fi%
    }%
    \expandafter\docsvlist\expandafter{\discard@pages}
  }%
}
\newif\ifkeeppage
\newcommand{\keeppages}[1]{
  \xdef\keep@pages{#1}
  \AtBeginShipout{
    \keeppagefalse%
    \renewcommand*{\do}[1]{
      \ifnum\value{page}=##1\relax%
        \keeppagetrue
        \gdef\do####1{}
      \fi%
    }%
    \expandafter\docsvlist\expandafter{\keep@pages}
    \ifkeeppage\else\AtBeginShipoutDiscard\fi
  }%
}

\newtheorem{Theorem}{Theorem}

\newtheorem{Definition}{Definition}

\ifCLASSINFOpdf

\begin{document}
%


\title{Non-Linear Analog Processing Gains in Task-Based Quantization}
\author{%
  \IEEEauthorblockN{Marian Temprana Alonso\IEEEauthorrefmark{1},
                    Farhad Shirani\IEEEauthorrefmark{1},
                    Neil Irwin Bernardo\IEEEauthorrefmark{2},
                    Yonina C. Eldar\IEEEauthorrefmark{3}
                   }
  \\\IEEEauthorblockA{\IEEEauthorrefmark{1}%
                   School of Computing  and Information Sciences,
                   Florida International University, Miami, FL,
                     \{mtempran,fshirani\}@fiu.edu}
  \IEEEauthorblockA{\IEEEauthorrefmark{2}%
                     Electrical and Electronics Engineering Institute,
                    University of the Philippines Diliman,
                    Quezon City, Philippines,
     neil.bernardo@eee.upd.edu.ph}
    \IEEEauthorblockA{\IEEEauthorrefmark{3}
    Department of Mathematics  and Computer Science, Weizmann Institute of Science, \\Rehovot, Israel,
yonina.eldar@weizmann.ac.il
    }
}

\maketitle

\begin{abstract}
In task-based quantization, a multivariate analog signal is transformed into a digital signal using a limited number of low-resolution analog-to-digital converters (ADCs). This process aims to minimize a fidelity criterion, which is assessed against an unobserved task variable that is correlated with the analog signal. The scenario models various applications of interest such as channel estimation, medical imaging applications, and object localization. This work explores the integration of analog processing components—such as analog delay elements, polynomial operators, and envelope detectors—prior to ADC quantization.
 Specifically, four scenarios, involving different collections of analog processing operators are considered:  (i) arbitrary polynomial operators with analog delay elements, (ii) limited-degree polynomial operators, excluding delay elements, (iii) sequences of envelope detectors, and (iv) a combination of analog delay elements and linear combiners. For each scenario, the minimum achievable distortion is quantified through derivation of computable expressions in various statistical settings.   
 It is shown that analog processing can significantly reduce the distortion in task reconstruction. Numerical simulations in a Gaussian example are provided to give further insights into the aforementioned analog processing gains. 

\end{abstract}
 
\section{Introduction}

Sensing, communication, and data compression systems utilize analog-to-digital converters (ADCs) to transform observed continuous-time analog signals into digital signals which can then be efficiently processed, communicated, and stored \cite{BR,ADCpower,chi2017subspace,corey2017wideband,shlezinger2019hardware,Salamatian:2019,Xi:2021,Khalili:2021,bernardo2022,shirani2022quantifying, shirani2022mimo,eldar2015sampling}. An ADC typically samples the signal at equally-spaced time intervals, and the amplitude of each sample is sequentially mapped onto a finite collection of quantization bins via comparison with pre-determined thresholds. The number of quantization bins is determined by the resolution of the ADC, and is quantified in terms of its output bits, e.g., a one-bit ADC has two quantization bins and its operation is parameterized by a single ADC threshold. Increasing the ADC resolution leads to reduced distortion. However, the ADC power consumption grows exponentially in the number of output bits. More precisely, in theory, the power consumption of an ADC is proportional to $f_s2^{n_q}$, where $f_s$ is the sampling rate and $n_q$ is the number of output bits of the ADC \cite{BR, walden1999analog}. As an example, the power consumption of current commercial
high-speed ($\geq$ 20 GSample/s), high-resolution
(e.g., 8-12 bits) ADCs is around 500 mW per ADC \cite{zhang2018low}.
This has led to significant recent interest in the use of low-resolution ADCs in data acquisition and processing systems and the design of hardware architectures and algorithms which mitigate the resulting loss in distortion due to coarse quantization.

Task-based quantization has emerged as a promising solution to mitigate the aforementioned rate-loss due to coarse quantization using low resolution ADCs  \cite{shlezinger2019hardware,bernardo2023analysis,Salamatian:2019,Xi:2021,Zirtiloglu:2022,Rini:2017,bernardo2021sep,Khalili:2021,Li:2021,bernardo2022}. 
The idea in task-based quantization is that the analog signal observed by the system is often digitized to be processed towards accomplishing a specific task, e.g., channel estimation, object localization, or pattern recognition in medical imaging \cite{shlezinger2021deep,shlezinger2019hardware,malak2023hardware,shohat2019deep}. Consequently, the ADCs and their accompanying analog processing circuits may be designed in a way to extract the task-relevant bits of information from the analog signal, while filtering out the irrelevant information through the lossy quantization process. In other words, the analog processing components and ADC thresholds are designed so that the distortion between the task reconstruction and the ground-truth task is minimized, rather than minimizing the distortion between the original signal and its reconstruction in the digital domain  \cite{shlezinger2019hardware,Salamatian:2019,bernardo2023}. 
Consequently, performance gains in task-based quantization are achieved by employing a hybrid analog/digital (A/D) architecture and jointly designing the analog pre-quantization mapping and digital post-quantization mapping with respect to the underlying task.

Prior design frameworks for task-based quantization have focused on linear processing in the analog domain. In this work, we consider the use of non-linear analog processing operators using implementable collections of analog components --- consisting of analog delay elements, polynomial operators, and envelope detectors prior to ADC quantization --- to further mitigate the coarse quantization distortion loss when using low resolution ADCs. This builds upon recent works \cite{shirani2022quantifying, shirani2022mimo,alonso2022capacity}, where the design and implementation of such circuit components for high frequency applications was considered in the context of wireless communications. It was shown that the power consumption of these analog processing components is negligible compared to that of the ADCs, hence justifying their application in such scenarios. Particularly, we consider four scenarios using analog operators consisting of: (i) arbitrary polynomial operators with analog delay elements, (ii) limited-degree polynomial operators, excluding delay elements, (iii) sequences of envelope detectors, and (iv) a combination of analog delay elements and linear combiners. In each scenario, we quantify the fundamental performance limits, in terms of achievable distortion in task reconstruction under general statistical assumptions on the task statistics. 
Furthermore, given a fixed ADC power budget --- using a fixed number and resolution of ADCs --- we show that the resulting task-reconstruction distortion decreases compared to the prior approach of using linear analog processing.  



\noindent {\em Notation:}
 The set $\{1,2,\cdots, n\}, n\in \mathbb{N}$ is represented by $[n]$. 
The vector $(x_1,x_2,\hdots, x_n)$ is written as $x(1\!\!:\!\!n)$ and $x^n$, interchangeably. 
The $i$th element is written as $x(i)$ and $x_i$, interchangeably. 
An $n\times m$ matrix is written as $h(1\!\!:\!\!n,1\!\!:\!\!m)=[h_{i,j}]_{i,j\in [n]\times [m]}$.
Sets are denoted by calligraphic letters such as $\mathcal{X}$.

\section{Problem Formulation}
\label{sec:formulation}
\begin{figure*}[!t] 
  \centering
  \includegraphics[width=0.95\textwidth]{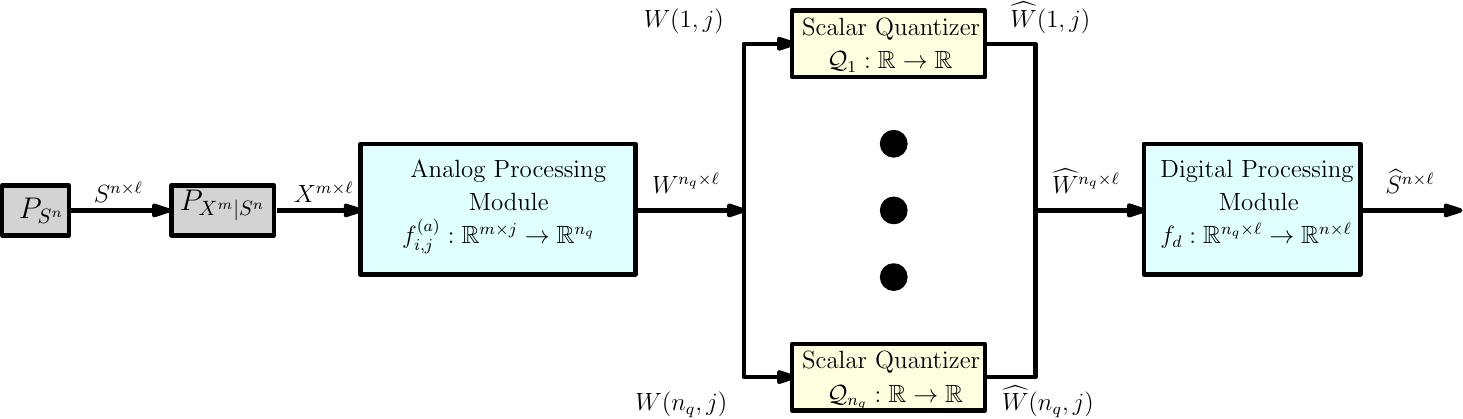}
  \caption{The task-based quantization setup.}
  \label{fig:overview}
  \vspace{-.15in}
\end{figure*} 
The task-based quantization setting considered in this work is shown in Figure \ref{fig:overview}.
In the following, we describe the general problem formulation, and provide examples in the context of channel estimation as a motivating application. 

\noindent \textbf{Task vector:} The (unobserved) sequence of {task vectors}  \( {S}^{n\times \ell}= (S^n(1),S^n(2),\cdots,S^n(\ell))\) are independently and identically distributed according to an underlying probability distribution $P_{S^n}(\cdot)$ defined on $\mathbb{R}^{n\times \ell}$, where \( n\in \mathbb{N} \) is the dimension of the task vector and $\ell\in \mathbb{N}$ is the blocklength. The vector $S^n(j)$ is the task vector at \textit{time} $j$, $j\in [\ell]$. The objective in task-based quantization is to produce an accurate reconstruction of the task vector based on a sequence of coarsely quantized indirect observations. As an example, in the context of channel estimation, the task vector  $S^n(j)$ represents the channel coefficient matrix at time $j$, and the objective is to produce an accurate channel estimate via indirect observations acquired by sending a sequence of pilot signals over the channel. 

\noindent \textbf{Measurement Vector:} The (observed) sequence of measurements is a sequence of real-valued vectors $X^{m\times \ell}=(X^m(1),X^m(2),\cdots,X^m(\ell)$, where $m\in \mathbb{N}$. Each $X^m(j),j\in [\ell]$ is produced conditioned on the realization of the task-vector $s^n(j)$ according to the conditional distribution $P_{X^m|S^n}(\cdot|s^n(j))$. For instance, in the context of channel estimation, the measurement vector at time $j$ models the analog channel output when a pilot signal is sent over the channel. 

\noindent \textbf{Analog Processing Functions:} The measurement vectors $X^{m\times \ell}$ are fed sequentially to a collection of analog processing functions $f^{(a)}_{i,j}: \mathbb{R}^{m\times j}\to \mathbb{R}, i\in [n_q], j\in [\ell]$, where $n_q\in \mathbb{N}$, and the choice of  $f^{(a)}_{i,j}, i\in [n_q],j\in [\ell]$ is restricted by the limitations of the analog circuit design as discussed in the sequel. In general, we assume that the analog processing functions at time $j$ are chosen from a set $\mathcal{F}_{a,j}$ of \textit{implementable analog functions.}
The output of the analog processing functions is denoted by $W^{n_q\times \ell}$, where
$W_{i,j}\triangleq f^{(a)}_{i,j}(X^{m\times i}), i\in [n_q], j\in [\ell]$,  $f^{(a)}_{i,j} \in \mathcal{F}_{a,j}$, and $n_q\in \mathbb{N}$. Note that in the general scenario described here, $W_{i,j}$ may casually depend on the past realizations of the measurement vector. The analog processing functions may consist of linear combiners, delay elements, non-linear operators such as low degree polynomial operators, and envelope detectors \cite{shirani2022quantifying,khalili2021mimo,khalili2018mimo,shlezinger2019hardware}. For a fixed number and resolution of ADCs, our objective is to quantify the gains due to the use of each of the aforementioned classes of non-linear analog processing functions, in terms of achievable distortion, in comparison with linear analog processing. 

\noindent\textbf{ADC Module.} At time $j\in [\ell]$, the processed signal vector $W^{n_q}(j)$ is fed to a set of $n_q$ ADCs each with $\kappa\in \mathbb{N}$ output levels. The quantization output is defined as $\widehat{W}^{n_q}(j)$, where
\begin{align}
\label{eq:ADC}
\widehat{W}(j,k)=
k \iff  W(j,i)\in [t_j(i,k),t_j(i,k+1)],
\end{align}
 $k\in [0,\kappa-1]$, $i\in [n_q]$, 
and we have defined $t_j(i,0)\triangleq -\infty$ and  $t_j(i,\kappa)\triangleq\infty$. We call $t_j^{n_q\times (\kappa)}$ the \textit{threshold matrix} at time $j$. $\widehat{W}(j,i)$ is called the quantization output of the $\kappa$-level ADC with thresholds $t_j^{n_q}(i)$ for input $W_j(i)$. 

\noindent \textbf{Digital Processing Function:} A digital processing function $f_d:\mathbb{R}^{n_q\times \ell}\to \mathbb{R}^{n\times \ell}$ acts on the sequence of quantized  vectors  $\widehat{W}^{n_q\times \ell}$ to produce the task reconstruction $\widehat{S}^{n\times \ell}$. There are no restrictions on the choice of the digital processing function.

\noindent \textbf{Distortion Function:} Given  $d:\mathbb{R}^n\times \mathbb{R}^n\to \mathbb{R}^{+}$, the  $\ell$-shot distortion is defined as:
    \vspace{-0.1in}
\begin{align*}
    d_{\ell}\triangleq \frac{1}{\ell}\sum_{j=1}^{\ell}\mathbb{E}_{S^{n\times \ell}, X^{m\times \ell}}(d(S^n(j),\widehat{S}^n(j))).
    \vspace{-0.1in}
\end{align*}
In summary, a task-based quantization setup is characterized by the tuple $(n,m,P_{S^n}, P_{X^m|S^n}, (\mathcal{F}_{a,j})_{j\in \mathbb{N}},n_q,\kappa,d(\cdot,\cdot))$. 

Given a collection of analog processing functions $(f^{(a)}_{i,j})_{i\in [n_q],j\in [\ell]}$ and thresholds $t_j^{n_q\times \kappa},j \in [\ell]$, the digital processing function minimizing distortion is given by:
    \vspace{-0.1in}
\begin{align*}
f^*_d=\argmin_{f_d:\mathbb{R}^{n_q\times \ell}\to \mathbb{R}^{n\times \ell}} \mathbb{E}_{S^{n\times \ell},X^{m\times \ell}}(d(f_d(W^{n_q\times \ell}),S^{n\times \ell})).
\end{align*}
For instance, if $d(\cdot,\cdot)$ is the square error distortion function, then by the orthogonality principle, we have $f^*_d(W^{n_q\times \ell})=\mathbb{E}({S}^{n\times \ell}|W^{n_q\times \ell})$. Since there are no restrictions on the choice of the digital processing functions, in the sequel, we always assume that the optimal digital processing function is used for reconstruction, i.e., $\widehat{S}^{n\times \ell}=f^*_d(W^{n_q\times \ell})$. Consequently, we focus on the optimization problem for the choice of analog processing functions and quantization thresholds.

\noindent \textbf{System Objective:} The objective in task-based quantization is to find the optimal choice of system parameters which minimize the achievable distortion given a fixed number and resolution of ADCs and a fixed collection of implementable analog processing functions $\mathcal{F}_{a,j},j\in \mathbb{N}$, 
To elaborate, the minimum $\ell$-shot achievable distortion is defined as:
    \vspace{-0.08in}
\begin{align}
\label{eq:obj}
    d^*_{\ell}\triangleq \min_{\substack{(f^{(a)}_{i,j})_{i\in [n_q],j\in [\ell]}\in \mathcal{F}_{a,j}\\ t_j^{n_q\times \kappa}\in \mathbb{R}^{n_q\times \kappa}}}\frac{1}{\ell}\sum_{j=1}^{\ell}\mathbb{E}_{S^{n\times \ell}, X^{m\times \ell}}(d(S^n(j),\widehat{S}^n(j))).
\end{align}
The collection of functions $(f^{(a)}_{i,j})_{i\in [n_q],j\in [\ell]}$ and thresholds $t_j^{n_q\times \kappa}, j\in [\ell]$ minimizing \eqref{eq:obj} are called the $\ell$-shot optimal functions and thresholds, respectively. Our objective is to characterize $ d^*_{\ell}$ and the corresponding processing functions and thresholds. 

\section{An Illustrative Gaussian Example}
\label{sec:example}
 In order to motivate the use of non-linear processing prior to quantization, in this section, we focus on a simple Gaussian example, and provide an intuitive justification of the performance gains due to using non-linear processing over linear processing. In the subsequent sections, we build upon the intuition provided by this example, and study the fundamental performance limits of the general task-based quantization problem using various classes of non-linear analog processing functions. Section \ref{sec:sim} numerically evaluates the achievable distortion in each of the scenarios considered in this section.
 
 Let us take $n=m=1$, and let the task be characterized by a zero-mean, unit variance, Gaussian random variable, i.e., $S\sim \mathcal{N}(0,1)$. Additionally, let us assume that the measurement vector is produced by passing the task through a Gaussian additive channel, i.e., $X=S+N, N\sim \mathcal{N}(0,\sigma_N^2)$, where $\sigma_N\in \mathbb{R}$ and $S$ and $N$ are independent of each other. Furthermore, let the quantization system be equipped by two one-bit ADCs, i.e., $n_q=\kappa=2$. Finally, we take $d(s,s')= (s-s')^2$ as the square distortion. We consider four scenarios, and find the minimum
achievable distortion in each case.
\subsubsection{Scenario 1. Linear Analog Processing} In this scenario, we restrict $f_a:\mathbb{R}\to \mathbb{R}$ to affine transformations, i.e., $\mathcal{F}_{a,j}=\{f_{a}|f_a(x)= bx+c, b,c\in \mathbb{R}\}, j \in [\ell]$. It is straightforward to see, using the orthogonality principle, that the minimum one-shot achievable distortion is given by:
\begin{align}
\label{eq:lin}
    d^*_{1,lin}=\min_{ \tau_1,\tau_2: \tau_1<\tau_2}\mathbb{E}_{S,N}((S-\widehat{S})^2),
\end{align}
where $\widehat{S}\triangleq \mathbb{E}(S|\widehat{X}_{\tau_1,\tau_2})$ and $\widehat{X}_{\tau_1,\tau_2}$ is the quantization output of a three-level ADC with thresholds $(-\infty,\tau_1,\tau_2,\infty)$ for input $X$ (see Equation \eqref{eq:ADC}). 
\subsubsection{Scenario 2. Quadratic Analog Operators} In this scenario, we choose $f_a:\mathbb{R}\to \mathbb{R}$ from the set of all quadratic functions, i.e., $\mathcal{F}_{a,j}= \{f:\mathbb{R}\to \mathbb{R}| f(x)= ax^2+bx+c, a,b,c\in \mathbb{R}\}, j \in [\ell]$. Let $\tau_1\leq \tau_2\leq \tau_3$ be arbitrarily chosen real numbers. Define $f_{1}(x)= (x-\tau_1)(x-\tau_3)$ and $f_2(x)= (x-\tau_2)$.
In this case, $W_{1}= (X-\tau_1)(X-\tau_3)$ and $W_2=X-\tau_2$ are the ADC inputs. We set the ADC thresholds to zero, so that $\widehat{W}_1= \mathbbm{1}(X\in [\tau_1,\tau_3])$ and $\widehat{W}_2=\mathbbm{1}(X\in [\tau_2,\infty])$. Thus, receiving $\widehat{W}_1$ and $\widehat{W}_2$ is equivalent to receiving the quantization output for quantizing $X$ with a four-level ADC with thresholds $\tau_1,\tau_2,\tau_3$. Consequently, 
\begin{align}
\label{eq:non}
    &d^*_{1,quad}=\min_{ \tau_1,\tau_2,\tau_3: \tau_1<\tau_2<\tau_3}\mathbb{E}_{S,N}((S-\widehat{S})^2),
\end{align}
where $\widehat{S}=\mathbb{E}(S|\widehat{X}_{\tau_1,\tau_2,\tau_3})$ and $\widehat{X}_{\tau_1,\tau_2,\tau_3}$ is the quantization output of a four-level ADC with thresholds $(-\infty,\tau_1,\tau_2,\tau_3,\infty)$ for input $X$. Note that this is an improvement over the achievable distortion of Scenario 1. In fact, to achieve $d^*_{1,quad}$ using linear analog processing, one needs to use three one-bit ADCs instead of two one-bit ADCs, thus requiring a fifty percent increase in  ADC power consumption.
\subsubsection{Scenario 3. Envelope Detectors} In this scenario, we assume the quantization system is equipped with envelope detectors, which can perform absolute value operations on the analog signal. Let $\tau_1\leq \tau_2\leq \tau_3$ be arbitrarily chosen real numbers. Define $f_1(x)= |x-\frac{\tau_1+\tau_3}{2}|$ and $f_2(x)= x$. Furthermore, let the ADC thresholds be $t_1= \frac{\tau_3-\tau_1}{2}$ and $t_2= \tau_2$. Then, 
\begin{align*}
    &\widehat{W}_1= \mathbbm{1}(|X- \frac{\tau_1+\tau_3}{2}|<\frac{\tau_3-\tau_1}{2})=\mathbbm{1}(X\in [\tau_1,\tau_3]),\qquad \widehat{W}_2= \mathbbm{1}(X>\tau_2).
    \end{align*}
Consequently, the achievable distortion is equal to that of Scenario 2, and improves the distortion in Scenario 1. In general the use of polynomial operators (Scenario 2) leads to lower achievable distortion compared to envelope detectors (Scenario 3), however the circuit design of envelope detectors is more straightforward than that of polynomial operators \cite{alonso2022capacity}, hence there is a trade-off between design complexity and achievable distortion between these two scenarios.

 It can be noted that in Scenarios 1-3, since $f_a$ is memoryless, and its output at time $j$ only depends on the input at time $j$, the minimum $\ell$-shot achievable distortion is equal to the minimum one-shot achievable distortion for all $\ell\in \mathbb{N}$. 
\subsubsection{Scenario 4. Analog Delay Elements} In this scenario, we consider the use of analog delay elements, which allows for causal memory in the analog processing functions. That is, we consider a processing function at time $j\in \mathbb{N}$ which is an affine function of the form $f_{a,j}:\mathbb{R}^j\to \mathbb{R}$ and $f_{a,j}$ takes $X^{m\times j}$ as input.  The two-shot  minimum achievable distortion is given by: 
\begin{align}
\label{eq:delay}
    d^*_{2,delay}= \min_{\tau_1,\tau_2,\tau_3,\tau_4,a_1,a_2}\frac{1}{2}\sum_{j=1}^2\mathbb{E}((S_j-\widehat{S}_j)^2),
\end{align}
 where $\widehat{S}_j= \mathbb{E}_{S_j|X_1,X_2}(S_j|\widehat{X}_1,\widehat{X}_2)$ and $\widehat{X}_1$ is the quantization output of a three-level ADC with thresholds $(-\infty,\tau_1,\tau_2,\infty)$ for input $X_1$ and $\widehat{X}_2$  is the quantization output of a three-level ADC with thresholds $(-\infty,\tau_3,\tau_4,\infty)$ for input $a_1X_1+a_2X_2$. It should be noted that the optimization in this scenario is over a larger search space compared to that of Scenario 1, as it allows for two-dimensional quantization, in the $(X_1,X_2)$ space rather than only the $X_2$ space, in the second time-slot. The optimization reduces to that of Scenario 1 by restricting to $a_1=0,a_2=1$. This achievable distortion is numerically evaluated in Section \ref{sec:sim}. We show that in this simple scenario, the gains due to the additional delay element are negligible compared to Scenario 1. However, if the use of delay elements is further augmented by analog polynomial operators, then we achieve significant gains over the previous three scenarios. 

\section{Fundamental Performance Limits in Task-Based Quantization}
\label{sec:performance_limits}
\subsection{Finite-degree Polynomial Operators and Delay Elements}
\label{sec:FD}
We consider a setup equipped with finite-degree polynomial operators with delay elements.  That is, we consider
the following set of implementable functions:
\begin{align*}
    \mathcal{F}^t_{a,j}= \{f(\cdot)| f(x^{m\times j})= &\sum_{\substack{(k_{u,v},u\in [m], v\in [j]):\\ \sum_{u,v}k_{u,v}\leq t}}
    b_{k^{m\times j}}\prod_{v\in [m], u\in [j]}x^{k_{u,v}},b_{k^{m\times j}}\in \mathbb{R}\},\\
&    \mathcal{F}_{a,j}= \cup_{t\in \mathbb{N}} \mathcal{F}^t_{a,j}, \quad j\in \mathbb{N} .
\end{align*}
\begin{Theorem}
\label{th:1}
    Consider a task-based quantization setup parametrized by  $(n,m,P_{S^n}, P_{X^m|S^n}, (\mathcal{F}_{a,j})_{j\in \mathbb{N}},n_q,\kappa,d(\cdot,\cdot))$ as described in the prequel. Assume that there exists $\mathbf{s}\in \mathbb{R}^{m}$ such that $\mathbb{E}(d(S^{m},\mathbf{s}))\leq \infty$. 
    The minimum achievable $\ell$-shot distortion for asymptotically large $\ell$ is given by:
    \begin{align}
    \label{eq:th:1}
    \lim_{\ell\to \infty}d^*_{\ell}= \min_{P_{\widehat{S}^m|X^n}:I(X^n;\widehat{S}^m)\leq n_q}\mathbb{E}_{S,X} (d(S^n,\widehat{S}^n)),
    \end{align}
    where $P_{S^m,X^n,\widehat{S}^m}\triangleq P_{S^m,X^n}P_{\widehat{S}^m|X^n}$, i.e., the Markov chain $S^n \leftrightarrow X^m\leftrightarrow \widehat{S}^n$ holds. 
\end{Theorem}
The distortion is then equal to the indirect distortion-rate function (iDRF) evaluated at compression rate  $n_q$ bits per input symbol. The proof follows by noting that using the multivariate Taylor expansion, any quantizer used for indirect source coding can be well-approximated, with arbitrary precision, using a finite-degree polynomial. Consequently, the optimal quantization scheme achieving the iDRF can be implemented using the analog processing functions, and its output (bits) can be passed through the ADCs without any further modification on the digital side. That is, the analog processing function is chosen such that its output is equal to that of the optimal compression function in the equivalent indirect source coding problem. Note that the output of the optimal compression function is binary, hence by setting the ADC thresholds equal to $\frac{1}{2}$, the binary analog processing outputs are recovered without further distortion on the digital side. 
The complete proof is given in Appendix \ref{App:th:1}. 


\subsection{Memoryless Finite-degree Polynomial Operators}
Implementing large analog delay elements may not be practically possible due to synchronization and chip space limitation issues. In this section, we consider a task-based quantization setup equipped with finite-degree polynomial operators without delay elements:
\begin{align*}
    \mathcal{F}^t_{a,j}= \{f(\cdot)| f(x^{m})= &\sum_{\substack{(k_{u},u\in [m]):\\ \sum_{u}k_{u}\leq t}}
    b_{k^{m}}\prod_{v\in [m]}x^{k_{u}},b_{k^{m}}\in \mathbb{R}\},j\in \mathbb{N}.
\end{align*}
Note that this can be considered as  the one-shot version of the scenario considered in Section \ref{sec:FD}.
\begin{Theorem}
\label{th:2}
    Consider a task-based quantization setup parameterized by  $(n,m,P_{S^n}, P_{X^m|S^n}, (\mathcal{F}_{a,j})_{j\in \mathbb{N}},n_q,\kappa,d(\cdot,\cdot))$. The minimum achievable distortion is given by:
    \begin{align}
    \label{eq:th:2}
    d^*_{\ell}= \min_{\substack{f:\mathbb{R}^m\to [\kappa^{n_q}]\\ g:[\kappa^{n_q}]\to \mathbb{R}^n}}\mathbb{E}_{S,X} (d(S^n,\widehat{S}^n)),
    \end{align} 
    for all $\ell\in \mathbb{N}$, where $\widehat{S}\triangleq g(f(X)))$.
\end{Theorem}
The proof follows by similar arguments as that of Theorem \ref{th:1}. We provide an outline in the following. We first note that since the system is not equipped with delay elements, the reconstruction at time $j$ only depends on the input at time $j$. Consequently, the $\ell$-shot minimum achievable distortion is the same for all values of $\ell\in  \mathbb{N}$. Hence, it suffices to consider the one-shot distortion. Furthermore, the ADCs can produce at most $\kappa^{n_q}$ Voronoi regions, which implies that the right-hand-side term in \eqref{eq:th:2} is a lower-bound for the achievable distortion. On the other hand, similar to the proof of Theorem \ref{th:1}, using the multi-variate version of Taylor's approximation, any quantizer with $\kappa^{n_q}$ Voronoi regions can be constructed using finite-degree polynomials and $n_q$ ADCs each with $\kappa$ quantization levels. This implies that the right-hand-side term in \eqref{eq:th:2} is an upper-bound for the achievable distortion. 

\subsection{Low-Degree Polynomials without Delay Elements}
It is shown in \cite{alonso2022capacity,shirani2022quantifying}, that although the power consumption of low-degree polynomial operators  such as quadratic operators may be significantly smaller than that of ADC components, the power consumption grows with polynomial degree, and becomes significant for high-degree polynomials. 
As a result, in this section we focus on the use of low degree polynomial operators with no delay elements. To derive computable, closed-form expressions for the achievable distortion, we focus on the scalar measurements and one-bit ADCs, i.e., $m=1, \kappa=2$. We consider the set of implementable analog functions $\mathcal{F}^{\delta}_{a,j}= \{f(\cdot)| f(x)= \sum_{i=0}^{\delta}a_i x^i, a_i\in \mathbb{R}\}$, where $\delta\in \mathbb{N}$ is the maximum polynomial degree. The following characterizes the minimum achievable distortion in this scenario. 
\begin{Theorem}
\label{th:3}
Consider a task-based quantization setup parameterized by  $(n,1,P_{S^n}, P_{X^m|S^n}, (\mathcal{F}^{\delta}_{a,j})_{j\in \mathbb{N}},n_q,2,d(\cdot,\cdot))$ as described in the prequel. Then, 
\begin{align*}
    d^*_{\ell}=\min_{\substack{(\tau_i)_{i\in [\Gamma]}\\ g:[\Gamma+1]\to \mathbb{R}^n}} \mathbb{E}_{S,X}(d(S^n,g(\widehat{X}))),
\end{align*}
where $\widehat{X}$ is the quantization output of a $(\Gamma+1)$-level ADC with thresholds $(-\infty, \tau_1,\tau_2,\cdots,\tau_{\Gamma}, \infty)$ and input $X$,  and 
\begin{align*}
& \Gamma\triangleq\min(2^{n_q}, \Gamma')
\qquad    \Gamma'\triangleq
    \begin{cases}
         n_q\delta+1 \qquad & \text{ if $\delta$ is odd},
        \\  n_q\delta & \text{otherwise}
    \end{cases}.
\end{align*}
\end{Theorem}
Note that since the polynomial operators may have a constant non-zero bias, we may assume without loss of generality that the ADCs have zero thresholds, and incorporate the thresholds into the polynomial bias. Then, the proof of the theorem follows by noting that the output of the ADC changes at the roots of the polynomial operator. Each polynomial operator of degree $\delta$ has at most $\delta$ distinct roots, and since there are $n_q$ operators, they may have at most $\delta n_q$ different roots. On the other hand, for even-degree polynomials, the value for asymptotically large negative and positive inputs are the same, hence the ADC output is equal for both. Consequently, there are at most $\Gamma'$ different quantization Voronoi regions as a result of the ADC operation. The complete proof is provided in Appendix \ref{App:th:2}. It should be noted that the result may be generalized to  $\kappa>2$ by using Proposition 4 in \cite{alonso2022capacity} to characterize the Voronoi regions.

\subsection{Envelope Detectors without Delay elements}
As shown in the circuit design and simulations of \cite{alonso2022capacity}, implementing envelope detectors to produce absolute value functions is less costly in terms of circuit design complexity and power consumption, compared to polynomial operators. Consequently, in this section, we consider
\begin{align*} 
\mathcal{F}^\delta_{a,j}=\{f(y)=A_s(x,b^s), x\in \mathbb{R}| s\in [\delta], 
    a^s\in \mathbb{R}^s\}, 
 \quad \delta\in \mathbb{N},
\end{align*}
where $A_1(x,b)\triangleq |x-b|, x,b\in \mathbb{R}$ and $A_s(x,b^s)\triangleq  A_1(A_{s-1}(x,b^{s-1}),b_s)= |A_{s-1}(x,b^{s-1})-b_s|, s\in \mathbb{N}$.
 That is, $\mathcal{F}^\delta_{env}$ consists of all functions which can be generated using sequences of $s\leq \delta$ concatenated envelope detectors with thresholds $b_1,b_2,\cdots,b_{s}$, respectively. 
 \begin{Definition}[\textbf{Fully-Symmetric Vector}] A vector $\mathbf{b}=(b_1,b_2,\cdots,b_{2^n})$ is called symmetric if $b_i+b_{2^n-i}=b_j+b_{2^n-j}, i,j\in [2^n-1]$. The vector
$\mathbf{b}$ is called fully-symmetric if it is symmetric 
and the vectors  $(b_1,b_2,\cdots,b_{2^{n-1}})$ and $(b_{2^{n-1}+1},b_{2^{n-1}+2},\cdots,b_{2^{n}})$ are both fully-symmetric for $n>2$ and symmetric for $n=2$.  
\end{Definition}

 \begin{Theorem}
Consider a task-based quantization setup parameterized by  $(n,1,P_{S^n}, P_{X^m|S^n}, (\mathcal{F}^{\delta}_{a,j})_{j\in \mathbb{N}},n_q,2,d(\cdot,\cdot))$ as described in the prequel. Then, 
\begin{align*}
    d^*_{\ell}=\min_{\substack{(\tau_i)_{i\in [\Gamma]}\in \mathcal{S}\\ g:[\Gamma+1]\to \mathbb{R}^n}} \mathbb{E}_{S,X}(d(S^n,g(\widehat{X}))),
\end{align*}     
 where $\Gamma\triangleq \min(2^{n_q}, n_q2^\delta)$, and $\mathcal{S}$ consists of the set of all vectors of length $n_q2^{\delta}$, which can be partitioned into $n_q$ fully-symmetric subvectors, each of length $2^{\delta}$.    
 \end{Theorem}
The proof follows by similar arguments as that of Theorem \ref{th:3} and \cite[Proposition 5]{alonso2022capacity}. 

\section{Simulation Results}
\label{sec:sim}
\begin{figure}[t]
    \centering
    \includegraphics[width=0.4\textwidth]{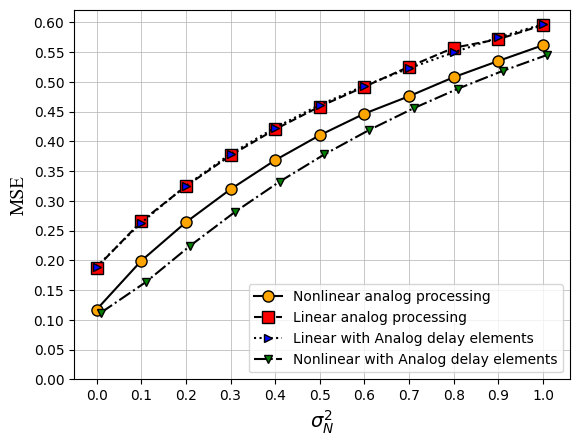}
    \vspace{-0.15in}
    \caption{Comparison of MSE distortion for linear and nonlinear analog processing with and without analog delay elements for a jointly Gaussian scalar task  $S$ and measurement $X$, with $n_q=\kappa=2$.}
    \label{fig:simulations}
    \vspace{-0.2in}
\end{figure}

 Let us consider the task-based quantization setup considered in Section \ref{sec:example}. Figure \ref{fig:simulations} provides a numerical evaluation of the achievable distortion in this setup under each of the scenarios considered in Section \ref{sec:performance_limits}. The linear analog processing plot (red square markers) shows the achievable distortion when only linear analog processing is used without delay element, and it serves as a baseline for the other schemes. It is derived by evaluating Equation \eqref{eq:lin} and sweeping over all possible values of $\tau_1,\tau_2$ with step-size 0.01. The linear processing with delay elements plot (blue triangle markers) is derived by evaluating Equation \eqref{eq:delay} by sweeping over values of $\tau_1,\tau_2,\tau_3,\tau_4,a_1,a_2$. It can be observed that in this simple scenario, the use of a single delay element while restricting to linear processing does not lead to a tangible performance improvement. The non-linear analog processing plot (orange circle markers) shows the achievable performance when quadratic polynomial operators are used without delay elements. It is derived by optimizing Equation \eqref{eq:non}. It can be observed that the use of quadratic operators improves the achievable distortion over the baseline. Lastly, the non-linear analog processing with delay elements plot (green triangle markers) shows the performance when polynomial operators with arbitrary degree and arbitrary number of delay elements can be used. It is derived by optimizing Equation \eqref{eq:th:1}. This serves as an outer-bound for the achievable distortion in the previously mentioned scenarios as it considers the most general subset of implementable analog functions.

\section{Conclusions}
\label{sec:conclusion}
The use of non-linear analog processing prior to quantization using low resolution ADCs in the task based quantization problem was studied. Several classes of non-linear analog processors were considered including analog delay elements, polynomial operators, and envelope detectors. In each scenario, the minimum achievable distortion was characterized and it was shown that the use of non-linear processing improves the achievable distortion. Simulations of a Gaussian task-based quantization setup were provided to illustrate these gains. 
\bibliographystyle{IEEEtran}
\bibliography{References}
\newpage
\begin{appendices}
\section{Proof of Theorem \ref{th:1}}
\label{App:th:1}
The proof builds upon previously known results in  the indirect source coding (IDS) setting (e.g., \cite{kipnis2015indirect,witsenhausen1980indirect}). Consequently,  we first provide a brief overview of these results. Consider an IDS scenario, where a source sequence $S^{n\time \ell}$ is generated IID according to $P_{S^n}$, and the observed variable $X^{m\times \ell}$ is generated by passing $S^n(j), j\in [\ell]$ through the memoryless channel $P_{X^m|S^n}$. The encoder uses an encoding function $e_{IDS}:\mathbb{R}^{m\times \ell}\to \{0,1\}^{\ell'}$ to produce $U^{\ell'}= e_{IDS}(X^{m\times \ell})$. The decoder uses the decoding function $h_{IDS}:\{0,1\}^{\ell'}\to \mathbb{R}^{n\times \ell}$ to produce the reconstruction $\widehat{S}^{m\times \ell}= h(U^{\ell'})$.
  The rate is defined as $R= \frac{\ell'}{\ell}$, and the distortion is defined as $D=\frac{1}{\ell}\sum_{j=1}^{\ell}d(S^{n}(j),\widehat{S}^{n}(j))$. It is known that the optimal rate-distortion tradeoff in the IDS scenario is characterized by the indirect rate-distortion function: \[R_{IDS}(D)=\inf_{\substack{P_{\widehat{S}^n|X^m}: \mathbb{E}(d(S^n,\widehat{S}^n))\leq D\\ S^n\leftrightarrow X^m \leftrightarrow \widehat{S}^n}} I(X^m;\widehat{S}^n),\quad D\geq 0.\]
Consequently, the minimum distortion $D$ achievable for a given rate $R$ is characterized by the distortion-rate function:
\[
D_{IDS}(R)= \inf_{\substack{P_{\widehat{S}^n|X^m}:I(X^m;\widehat{S}^n)\leq R\\ S^n\leftrightarrow X^m \leftrightarrow \widehat{S}^n}} \mathbb{E}(d(S^n,\widehat{S}^n)),\quad R\geq 0.
\]
We argue that $\lim_{\ell\to \infty} d^*_{\ell}= D_{IDS}(n_q)$. An outline of the achievability and converse proofs is given below. 
\\\textit{Proof of Achievability:} The proof builds upon the ideas introduced in the proof of \cite{shirani2022mimo}. 
Fix $\epsilon>0$, and let $\ell_{\epsilon},\ell'_{\epsilon}\in \mathbb{R}$ $e_{IDS}: \mathbb{R}^{m\times \ell_{\epsilon}}\to \{0,1\}^{\ell'_{\epsilon}}$ and $ h_{IDS}: \{0,1\}^{\ell'_{\epsilon}}\to \mathbb{R}^{m\times \ell_{\epsilon}}$ be such that $\frac{\ell'_{\epsilon}}{\ell_{\epsilon}}\leq n_q$ and $\frac{1}{\ell_{\epsilon}}\sum_{j=1}^{\ell_{\epsilon}}d(S^{n}(j),\widehat{S}^{n}(j))\leq D_{IDS}(n_q)+\epsilon$. Define the partition of $\mathbb{R}^{m\times \ell_{\epsilon}}$ corresponding to $e_{IDS}$ as $\mathsf{P}=\{\mathcal{P}_{j}|j \in [2^{\ell'_{\epsilon}}]\}$, where each $\mathcal{P}_{j}$ is equal to $e_{IDS}^{-1}(u^{\ell'_{\epsilon}})$ for some $u^{\ell'_{\epsilon}}\in \{0,1\}^{\ell'_{\epsilon}}$, i.e., $\mathsf{P}$ is the set of Voronoi regions formed by the quantization operation performed by $e_{IDS}$. 

We define a task-based quantization scheme by choosing the analog and digital processing functions according to $e_{IDS}(\cdot)$ and $h_{IDS}(\cdot)$ to achieve the desired distortion. To construct the analog processing functions, let us 
first define the collection of functions $
    f'_t(x^{m\times \ell_{\epsilon}})= (-1)^{mod_{2^t}(k)}||x^{m\times \ell_{\epsilon}}-\partial\mathcal{P}_k||_2$, where $x^{m\times \ell_{\epsilon}}\in \mathbb{R}^{m\times \ell_{\epsilon}}$, $t\in \{0,1,\cdots,{\ell'_{\epsilon}}-1\}$, $mod_b (a)$ denotes $a$ modulo $b$, $k\in [2^{\ell'_{\epsilon}}]$ is the index of the partition region for which $x^{m\times \ell_{\epsilon}}\in \mathcal{P}_k$,  and  $||x^{m\times \ell_{\epsilon}}-\partial\mathcal{P}_k||_2$ is the $\ell_2$ distance between $x^{m\times \ell_{\epsilon}}$ and the boundary of the region $\mathcal{P}_k$. By construction, the function $f'_t(\cdot)$ is continuous and its roots are the boundary points of the partition regions $\mathcal{P}_{k'}, k'\in [2^{\ell'_{\epsilon}}]$. Furthermore, its value is positive for all interior points of regions $\mathcal{P}_{k'}, k'\in [2^{\ell'_{\epsilon}}]$ for which $mod_{2^t}k'$ is even and is negative otherwise. As a result, $Sign(f_t(x^{m\times \ell_{\epsilon}})), t\in \{0,1,\cdots,{\ell'_{\epsilon}}-1\}$ is the binary representation of the index of $\mathcal{P}_k$, where $x^{m\times \ell_{\epsilon}}\in \mathcal{P}_k$, i.e. $Sign(f'_t(x^{m\times \ell_{\epsilon}}))=e_{IDS}(x^{m\times \ell_{\epsilon}})$. 
    We define the (polynomial) analog processing functions such that they produce the outputs of $f'_t(\cdot)$, so that the ADC output is $e_{IDS}(x^{m\times \ell_{\epsilon}})$. To this end, let $f^{\delta}_{t}(\cdot), t\in \{0,1,\cdots,\ell'_{\epsilon}-1\}$ be the best polynomial approximation of  $f'_{t}(\cdot)$ in terms of $\ell_{\infty}$ distance. It is well-known that $f_t^{\delta}\to f'_t$ as $\delta\to \infty$, and convergence is uniform over any compact subset of $\mathbb{R}^{m\times \ell_{\epsilon}}$. 

    We use the polynomial processing functions $f^{\delta}_{t}(\cdot), t\in \{0,1,\cdots,{\ell'_{\epsilon}}-1\}$ as follows. Let $\ell=\alpha\ell_{\epsilon}$ for some $\alpha\in \mathbb{N}$. The scheme uses a bank of $2m\times \ell_{\epsilon}$ delay elements as in \cite{khalili2021mimo}. That is, in the fist $\ell_{\epsilon}$ time-slots, the observed vectors $X^{m}(j), j\in [\ell_{\epsilon}]$ are input into a series of $m\times \ell_{\epsilon}$ analog delay elements. At each of the subsequent time-slots, $j\in [\ell_{\epsilon}+1,2\ell_{\epsilon}]$, the content of these delay elements are input to $f_{a,t}= f^{\delta}_{n_q j+ i}(\cdot), i\in \{0,1,\cdots,n_q-1\}$, where we have used the fact that $\ell'_\epsilon=n_q\ell_\epsilon$. The observed vectors $X^{m}(j), j\in [\ell_{\epsilon}+1,2\ell_{\epsilon}]$ are placed in the second set of $m\times \ell_{\epsilon}$ delay elements. These signals are processed in the next $\ell_{\epsilon}$ time-slots and this is repeated until the end of $\ell$th time-slot. The ADC thresholds are set to zero for all time-slots. The decoder uses the function $h_{IDS}$ for reconstruction.
    Consequently, in time-slots $[\ell_{\epsilon}+1,2\ell_{\epsilon}]$, the ADCs produce $U^{\ell'_{\epsilon}}=e_{IDS}(X^{m\times {\ell_{\epsilon}}})$. It follows that at time $\alpha \ell_{\epsilon}$, the decoder reconstructs $(\alpha-1)\ell_{\epsilon}$ reconstructions with expected distortion $\frac{\alpha-1}{\alpha}D_{IDS}(R)$. Note that each reconstruction is delayed by $\ell_{\epsilon}$ time-slots since we have skipped reconstruction in the first $\ell_{\epsilon}$ time-slots. 
    Consequently, for the last $\ell_{\epsilon}$ time-slots, the decoder outputs a fixed reconstruction $\mathbf{s}$ for all symbols. 
    The resulting distortion is $\frac{\alpha-1}{\alpha}D_{IDS}(R)+ \frac{1}{\alpha}\mathbb{E}(d(S^{m\times \ell_{\epsilon}},\mathbf{s}))$. This converges to $D_{IDS}(R)$ as $\alpha\to \infty$ and $\epsilon\to 0$ given that $\mathbb{E}(d(S^{m\times \ell_{\epsilon}},\mathbf{s}))$ is finite. 
    \\\textit{Proof of Converse:} Given a task-based quantization system with $(f^{(a)}_{i,j}, i\in [n_q], j\in [\ell])$ analog processing functions and $f_d:\mathbb{R}^{n_q\times \ell}\to \mathbb{R}^{n\times \ell}$ digital processing function, let us define $e_{IDS}:\mathbb{R}^{m\times \ell} \to \{0,1\}^{\ell\times n_q}$ such that $e_{IDS,i,j}(\cdot)\triangleq Sign(f^{(a)}_{i,j}(\cdot)), i\in [n_q], j\in [\ell]$,  where we have denoted $e_{IDS}(\cdot)= [e_{IDS,i,j}(\cdot)]_{i\in [n_q], j\in [\ell]}$, 
    and let us define $h_{IDS}(\cdot)\triangleq f_d(\cdot)$. By construction, the compression rate in this IDS system is $n_q$ bits per symbol, and it achieves distortion equal to that of the task-based quantization system, hence proving the  converse. \qed

\section{Proof of Theorem \ref{th:3}}
\label{App:th:2}
The proof builds upon \cite{shirani2022quantifying}, which characterizes surjective correspondence between the choice of
analog processing functions and the set of $(\Gamma+1)$-level ADCs. An outline of proofs for achievability and converse is given below.  
\\ First, we prove achievability. Let us fix an increasing sequence of thresholds $\tau_i\in \mathbb{R}, i\in [\Gamma]$, i.e., $\tau_i\leq \tau_{i+1}$ for all $i\in [\Gamma-1]$ and define $\tau_0=-\infty, \tau_{\Gamma+1}=\infty$. We wish to show that by appropriately choosing the analog processing functions $f^{(a)}_i(\cdot), i\in [n_q]$ and using zero-threshold one-bit ADCs, one can `emulate' a $\Gamma+1$-level ADC with  thresholds $\tau_i,i\in [\Gamma-1]$. 

Let us denote the collection of ADC outputs $Q(\cdot)\triangleq (Q_{1}(\cdot),Q_{2}(\cdot),\cdots,Q_{n_q}(\cdot))$, where $Q_{i}(x)\triangleq \mathbbm{1}(f^{(a)}_{a,i}(x)>0), i\in [n_q]$, and 
the functions $f^{(a)}_{i}(\cdot)$, $i\!\in\! [n_q]$ are polynomials of degree at most $\delta$. We call $Q(\cdot)$ \textit{the quantizer}. 
Define the associated partition of the quantizer $Q(\cdot)$ as $\mathsf{P}=\{\mathcal{P}_{\mathbf{i}}, \mathbf{i}\in [2^{n_q}]\}- \Phi$, where 
\[\mathcal{P}_\mathbf{i}= \{x\in\mathbb{R}| Q(x)= \mathbf{i}\}, \mathbf{i}\in [2^{n_q}].\] To prove achievability, 
it suffices to show that for any choice of $\tau_i, i\in [\Gamma]$, there exists a $Q(\cdot)$ such that $\mathsf{P}= \{[\tau_i,\tau_{i+1}), i\in [0,\Gamma]\}$. To this end, we need the output vector of the ADCs to change values at each $X=\tau_i$, and for the ADCs to output unique output vectors for each partition element. Note that any point of discontinuity of $Q(\cdot)$ is the root of the polynomial $f^{(a)}_{i}(x)$ for some $i\in [n_q]$.  Let $r^{(\ell-1) \delta n_q}$ be the sequence of roots of polynomials $f^{(a)}_{i}(\cdot), i\in [n_q]$ (including repeated roots), written in non-decreasing order, and let $\mathcal{C}=(\mathbf{c}_0,\mathbf{c}_1,\cdots, \mathbf{c}_{( \delta n_q})$ be the corresponding quantizer outputs, i.e. $\mathbf{c}_{t-1}= \lim_{x\to r_t^-}Q(x), t\in [\delta n_q]$ and $\mathbf{c}_{\delta n_q}=\lim_{x\to\infty}Q(x)$. Following the terminology introduced in \cite{shirani2022quantifying}, we call $\mathcal{C}$ the associated code of the quantizer. Note that $Q(\cdot)$ is completely characterized by its corresponding root sequence and associated code. From \cite[Proposition 4]{shirani2022quantifying}, it follows that for $\Gamma$ defined in the theorem statement, there exists an associated code $\mathcal{C}$ such that for any given root sequence $r^{\delta n_q}$, there exists a $Q(\cdot)$ with that root sequence and $\mathcal{C}$ as its associated code. Let $k_t, t\in \{1,\dots,\Gamma-1\}$ be the bit position which is different between $\mathbf{c}_{t-1}$ and $\mathbf{c}_{t}$. Consider a  quantizer $Q(\cdot)$ with associated polynomials $f^{(a)}_{i}(x)\triangleq -\prod_{t: k_t=i}(x-\tau_t), i\in [n_q]$. Then, $\tau_1,\tau_2,\cdots, \tau_{\gamma}$ are the non-decreasing sequence of roots of $f^{(a)}_{i}(\cdot), i\in [n_q]$, and the associated code of the quantizer $Q(\cdot)$ is $\mathcal{C}$ as desired. Furthermore, form Property 5) in \cite[Proposition 2]{shirani2022quantifying}, it follows that $f^{(a)}_{i}$ has degree at most $\delta$. This completes the proof.

Furthermore, from
\cite[Proposition 3]{shirani2022quantifying} it follows that for any 
choice of $\tau_i, i\in [\Gamma]$, there exists a root sequence for which the output of $Q(\cdot)$ is the same as that of a $\Gamma+1$-level ADC with quantization thresholds  $\tau_i, i\in [\Gamma]$.
This concludes the proof of achievability. 
\\The converse follows by noting from \cite[Proposition 2]{shirani2022quantifying} that $Q(\cdot)$ cannot produce more than $\Gamma$ distinct partition elements.

\end{appendices}

\end{document}